\documentclass[12pt]{iopart}

\usepackage{iopams}
\usepackage{graphicx}
\usepackage{bm}
\usepackage{colordvi,xcolor}
\newcommand{\dbar}{\mathchar'26\mkern-12mu d}

\begin{document}

\title{Work fluctuation theorems and free energy from kinetic theory }
\author{J. Javier Brey, M.J. Ruiz-Montero, and \'{A}lvaro Dom\'{\i}nguez}
\address{F\'{\i}sica Te\'{o}rica, Universidad de Sevilla,
Apartado de Correos 1065, E-41080, Sevilla, Spain}
\ead{brey@us.es}

\begin{abstract}
The formulation of the First and Second Principles of thermodynamics for a particle in contact with a heat bath and submitted to an external force is analyzed, by means of the  Boltzmann-Lorentz kinetic equation. The possible definitions of the thermodynamic quantities are discussed in the light of the H theorem verified by the distribution
of the particle. 
The  work fluctuation relations formulated by Bochkov and Kuzovlev, and by Jarzynski, respectively, are derived from the kinetic equation.  In addition, particle simulations using both the direct simulation Monte Carlo method and Molecular Dynamics, are used to investigate the practical accuracy of the results. Work distributions are also measured, and they turn out to be rather complex. On the other hand, they seem to depend very little, if any, on the interaction potential between the intruder and the bath.

\end{abstract}


\vspace{2pc}

%\submitto{\JSTAT}
\maketitle

\section{Introduction}
\label{s1}
There are in the literature several identities related with the work distribution associated to a process starting in a thermal equilibrium state \cite{ByK77,ByK81,Ja97a,Ja97b}. In particular, the so called Jarzynski fluctuation theorem or Jarzynski relation (JR) has been rederived in a variety of model systems \cite{Ja97b,Cr99,Mu03} and employed to discuss a series of  experiments \cite{Letal02,HyS01,Betal06}. On the other hand, some criticisms have been also raised about the correctness of the relation, concerning mainly the separation of the system from equilibrium along the process \cite{CyM04} and the definition of thermodynamic work used in the derivation
\cite{VyR08}. Although both criticisms were answered by Jarzynski and collaborators \cite{HyJ08,Ja04}, it is worth to consider them as well as the relevance  of work fluctuation theorems starting from a different level of description of the system. The analysis includes a work fluctuation relation by Bochkov and Kuzovlev (BK) \cite{ByK77,ByK81}, which will be shown to be closely related with the Jarzynski one.  The aim of this paper is to address the above issues as well as others related with the meaning and usefulness of the work relations, starting from a well established kinetic equation for a particle in contact with a heat bath.  

In thermodynamics, the free energy $F$ of an homogeneous and isotropic system at equilibrium is defined as
 \begin{equation}
 \label{1.1}
 F= U- TS,
 \end{equation}
 where $U$ is the {\em internal} energy, $T$ the absolute temperature, and $S$ the entropy. According with the
 Second Principle, the change of the free energy of a closed system in an infinitesimal quasistatic process is related to the 
 work $\dbar W$  performed by the system in the process by
 \begin{equation}
 \label{1.2}
 dF=-SdT- \dbar W.
 \end{equation}
 It follows that, for a finite quasistatic process carried out at constant temperature, the difference $ \Delta F$ between the final and initial equilibrium free energies is given by minus the total work $W_{T}$,
\begin{equation}
\label{1.3}
\Delta F= - W_{T}.
\end{equation}
In  equilibrium statistical mechanics,  the connection with thermodynamics for homogeneous  and isotropic systems is made through the relationship
\begin{equation}
\label{1.4}
F=-k_{B} T \ln Z.
\end{equation} 
Here $k_{B}$ is the Boltzmann constant and $Z$ the partition function of the system defined in the classical limit  as an integral over the phase space $\Gamma $ of the system,
\begin{equation}
\label{1.5}
Z \equiv \int d \Gamma\, e^{-H(\Gamma)/k_{B}T}\, ,
\end{equation}
with $H$ being the Hamiltonian of the system. A constant needed to render $Z$ dimensionless is omitted. If the Hamiltonian depends on a parameter, the free energy difference between two equilibrium states corresponding to two different values of the parameter can be obtained from the quasistatic work needed to go from one value to the other at constant temperature. Of course, the same difference can be formally computed by means of Eq. (\ref{1.4}).

Suppose a system initially at equilibrium with a temperature $T$, being $H_{0}(\Gamma)$ its Hamiltonian. Then, at $t=0$ the system  is submitted to a time dependent perturbation, $\phi(\Gamma,t)$ so that the Hamiltonian becomes $H(\Gamma,t)= H_{0}(\Gamma)+ \phi(\Gamma,t)$, with $\phi(\Gamma,0)=0$.  Along the process, the system remains isolated, i.e. there is no heat exchange with another system. Assume that the same process of variation of the Hamiltonian can be repeated many times, starting always  from the same macroscopic equilibrium state, and that the work $w(t)$ required in each individual process up to time $t$  is measured. Using the properties of the Liouville equation, Bochkov and Kuzovlev \cite{ByK77,ByK81} obtained the relation
\begin{equation}
\label{1.6}
 \langle e^{-w(t)/k_{B}T} \rangle =1,
 \end{equation}
for arbitrary $t>0$. The angular brackets denote an average over the ensemble of realizations of the process, i.e. over trajectories in phase space, and
\begin{equation}
\label{1.7}
w(t) \equiv -\int_{0}^{t} d\tau \sum_{i}{\bm v}_{i} (\tau) \cdot \frac{\partial \phi \left[\Gamma(\tau),\tau \right]}{\partial {\bm r}_{i} (\tau)}\, ,
\end{equation}
where the sum extends over all the particles in the system, $\Gamma(\tau)$ is the phase point obtained from $\Gamma$ due to the evolution of the system between $0$ and $\tau$. Similarly, ${\bm r}_{i}(\tau)$ and ${\bm v}_{i} (\tau)$ are the position and velocity of particle $i$ at time $\tau$, respectively. Notice that only the force associated with the perturbation,  which vanishes up to $t=0$,  is considered when evaluating this work. Also, let us emphasize that the work is defined with its usual sign in mechanics and not as in the thermodynamic relation given in Eq.\, (\ref{1.2}). Twenty years later, Jarzynski \cite{Ja97a, Ja97b} derived   for the same process the relation
\begin{equation}
\label{1.8}
\langle e^{-w^{\prime}(t)/k_{B}T} \rangle=  e^{- \Delta F(t) /k_{B}T} .
\end{equation}
In this expression, the angular brackets have the same meaning as in Eq.\ (\ref{1.6}), and $\Delta F \equiv F[T;H(t)]-F[T;H_{0}]$  is the free energy difference between the two equilibrium states corresponding to $H(\Gamma, t)$ and $H_{0}(\Gamma)$. 
It is important to realize that the system is  at equilibrium only at the initial time. As a consequence, the Jarzynski relation  provides a method to get the difference between equilibrium values of the free energy $F$ from measurements of the fluctuations of the work $w^{\prime}$ along  trajectories extending well inside non-equilibrium regions. The quantity $w^{\prime}$ is identified as the work performed during each repetition of the process. In spite of the difference between Eqs. (\ref{1.6}) and (\ref{1.8}), both results are mathematical identities, following directly from the Hamilton equations of motion and the form of the equilibrium canonical distribution. The apparent contradiction lies in the different definitions of work along a trajectory being used \cite{Ja07}. Jarzynski's expression is 
\begin{equation}
\label{1.9}
w^{\prime} (t) \equiv  \int_{0}^{t} d\tau \phi_{\tau} \left[ \Gamma(\tau), \tau \right],
\end{equation}
\begin{equation}
\label{1.10}
\phi_{\tau}(\Gamma,\tau) = \left( \frac{\partial \phi (\Gamma,\tau)}{\partial \tau} \right)_{\Gamma}\, .
\end{equation}
Again, the mechanical criterium for the sign of work has been used. Both work  fluctuation relations were originally derived by means of (reversible and deterministic) Hamiltonian dynamics, although later on they were proven to remain valid for Markov stochastic dynamics \cite{Ja97b}. A first question is whether the relations  also remain valid for irreversible non-equilibrium dynamics as provided by kinetic theory, not necessarily with an underlying Markov process. Another  significant issue is which are the right definitions of work and free energy to be used in the formulation of the Second Principle for these, in general, inhomogeneous systems, if one wants to keep the formulation given by Eq. (\ref{1.2}). A particularly relevant context in which to study the above points seems to be a small system in contact with a heat bath, which corresponds to an idealization of most of the reported experiments related with work fluctuation relations. It is fair to mention that some of the above issues, concerning stationary properties of inhomogeneous systems, have been extensively studied by means of density functional theory \cite{Ev92}.

The remaining part of the paper is organized as follows. In Sec. \ref{s2}, the Boltzmann-Lorentz (BL) kinetic equation for a particle in an external potential  and in contact with a heat bath is introduced and used to derive energy balance equations, pointing out the several options that appear when defining the thermodynamic energy and the work in a process. Also, a modification of the celebrated $H$ Boltzmann theorem is derived, leading to the identification of a thermodynamic potential that is associated with the free energy $F$ of the inhomogeneous system. Details of the proof are given in Appendix \ref{ap1}. The BK relation and the JR are derived from the kinetic equation in Sec.\ \ref{s3}. Both relations are explicitly checked by solving numerically the BL kinetic equation by means of the direct simulation Monte Carlo method in Sec. \ref{s4}. Equivalent results follow from Molecular Dynamics simulations in sufficiently dilute systems. In addition, the form of the work distributions along trajectories is investigated.  The last section of the paper contains a short summary and some final comments.

\section{Boltzmann-Lorentz kinetic equation  in the presence of an external field}
\label{s2}

To address the questions raised in the previous section, consider a particle (intruder) of mass $m$ immersed in a low density gas of particles of mass $m_{b}$ and number of particles density $n_{b}$. The gas is at equilibrium at temperature $T_{b}$, and it is assumed that the state of the gas is not affected by the state of the intruder, i.e. it acts as a thermal bath. There is an external force acting on the particle of the form
\begin{equation}
\label{2.1} 
{\bm F} = -\frac{\partial \phi ({\bf r},t)}{\partial {\bm r}}\, ,
\end{equation}
\begin{equation}
\label{2.2}
\phi({\bm r},t)= \phi_{0}({\bm r}) + \phi_{1}({\bm r},t),
\end{equation}
where $\phi_{1}({\bm r},t)$ vanishes for $t \leq 0$. The probability density $f({\bm r},{\bm v},t)$ of finding the particle at position ${\bm r}$ with velocity ${\bm v}$ at time $t$  obeys the Boltzmann-Lorentz (BL) equation \cite{RydL77}
\begin{equation}
\label{2.3}
 \frac{\partial f}{\partial t} + {\bm v} \cdot \frac{\partial f}{\partial {\bm r}}+ \frac{\bm F}{m} \cdot \frac{\partial f}{\partial {\bm v}}  = J_{BL} [{\bm r},{\bm v},t |f,f_{b}],
\end{equation}
with the BL  collision term given by 
\begin{eqnarray}
\label{2.4}
J_{BL} [{\bm r},{\bm v},t |f,f_{b}] =& \\\nonumber
 \int d{\bm v}_{1} \int d\Omega\,  \sigma (\Omega,g) g  \left[ f({\bm r},{\bm v}^{\prime},t) f_{b}(v^{\prime}_{1})-f({\bm r},{\bm v},t) f_{b}(v_{1}) \right]. &
\end{eqnarray}
Here ${\bm v}^{\prime}$  and $v^{\prime}_{1}$ denote the postcollisonal velocities, ${\bm g} \equiv {\bm v}-{\bm v}_{1}$ is the relative velocity of the intruder with respect to the gas particle before the collision, $\sigma$ is the differential cross section,  $d\Omega$ is the solid angle element, and the primes indicate post-collisional velocities. Moreover, $f_{b}(v_{1})$ is the (equilibrium) one-particle distribution function of the gas,
\begin{equation}
\label{2.5}
f_{b}(v_{1}) \equiv  n_{b} \varphi_{b} (v_{1}),
\end{equation}
\begin{equation}
\label{2.6}
\varphi_{b}(v_{1})=\left( \frac{m_{b}}{2 \pi k_{B} T_{b}} \right)^{3/2} e^{-m_{b}v^{2}_{1} /2 k_{B}T_{b}}  .
\end{equation}
The BL equation can be considered as an exact equation in the low density limit, if it is assumed that the gas acts as an equilibrium  bath with respect to the intruder, although the collisions between the intruder and the gas particles are left arbitrary, as long as they correspond to the qualitative picture of a repulsive part at short distances and a possible atractive part at larger distances, vanishing sufficiently fast in the limit of an infinite separation of the involved particles. In particular, let us emphasize that it does not presuppose anything about the macroscopic or thermodynamic description of the state of the particle. On the other side, it is assumed that the range of the interaction potential between the intruder and the bath particles is much shorter that the mean free path of the latter. As already mentioned, a particle inside an equilibrium fluid is the prototype of situations to which the work theorems have been applied, both in theoretical studies \cite{Ja07,HyJ09,Se12}, and in experiments \cite{HyS01,Letal02,Betal06,HSyK07}. The average kinetic energy of the intruder at time $t$ is
\begin{equation}
\label{2.7}
e(t) \equiv \int d{\bm r} \int d{\bm v} \frac{mv^{2}}{2}\, f({\bm r}, \bm v,t),
\end{equation}
and from Eq. (\ref{2.3}) it is obtained
\begin{equation}
\label{2.8}
\Delta e(t_{1},t_{2}) \equiv e(t_{2})-e(t_{1}) =Q(t_{1},t_{2})- W(t_{1},t_{2})
\end{equation}
with
\begin{equation}
\label{2.9}
W(t_{1},t_{2})= - \int_{t_{1}}^{t_{2}} dt \int d{\bm r} \int d{\bm v}\, {\bm v} \cdot {\bm F} f({\bm r},{\bm v},t) 
\end{equation}
and
\begin{equation}
\label{2.10}
Q(t_{1},t_{2}) = \int_{t_{1}}^{t_{2}} dt \int d{\bm r} \int d{\bm v}\, \frac{mv^{2}}{2} J_{BL} [{\bm r},{\bm v},t |f,f_{b}] .
\end{equation}
The physical meaning of the term denoted by $Q$, as representing the energy exchange with the gas bath through collisions, strongly suggests identifying it with the
heat dissipated in the process. Consistently, it seems appropriate to define $e$ as the {\em internal} energy of the intruder, and the term $W$ as the work, with the usual sign convention  in thermodynamics. Alternatively, the {\em total} energy average
\begin{equation}
\label{2.11}
e_{0}(t) \equiv \int d{\bm r} \int d{\bm v}\, \left[ \frac{m v^{2}}{2}\, + \phi ({\bm r},t) \right] f({\bm r},{\bm v},t)
\end{equation}
can be considered. Then, again from the BL equation one gets
\begin{equation}
\label{2.12}
\Delta e_{0}(t_{1},t_{2}) \equiv e_{0}(t_{2})-e_{0}(t_{1}) =Q(t_{1},t_{2})- W^{\prime}(t_{1},t_{2}),
\end{equation}
where $Q(t_{1},t_{2})$ is the same as in Eq.\ (\ref{2.10}) and
\begin{eqnarray}
\label{2-13}
W^{\prime}(t_{1},t_{2}) & = & - \int_{t_{1}}^{t_{2}} dt \int d{\bm r} \int d{\bm v}\ \frac{\partial \phi}{\partial t}\, f({\bm r},{\bm v},t) \nonumber \\
&=& - \int_{t_{1}}^{t_{2}} dt \int d{\bm r} \int d{\bm v}\ \frac{\partial \phi_{1}}{\partial t}\, f({\bm r},{\bm v},t).
\end{eqnarray}
Therefore, there is an apparent ambiguity in the definition of internal energy (and work), raising the issue of which of the two above definitions is consistent with the classical formulation of thermodynamics. Let us point out that in kinetic theory \cite{RydL77,McL89}, and also in usual hydrodynamics \cite{GyM62}, the local internal energy does not include the potential energy associated to an external field. In order to avoid misunderstandings, it is worth insisting on  that  the work expression considered by BK and given in Eq.\ (\ref{1.7}) does not correspond exactly to the work expression defined by Eq.\ (\ref{2.9}), since the force appearing in the former does not include the contribution from  the external potential acting already before $t=0$, i.e. the potential $\phi_{0}({\bm r})$. On the other hand, in the formulation of the JR, the difference between $\phi$ and $\phi_{1}$ disappears, since the difference, $ \phi_{0}$, does not depend on time.

Suppose for a while that the external field $\phi$ does not depend on time (e.g. $\phi_{1}=0$). Define the functional of the distribution function 
\begin{equation}
\label{2.14}
{\cal H}(t) \equiv \int d{\bm r} \int d{\bm v} f({\bm r},{\bm v}, t)  \left[  \ln f({\bm r},{\bm v},t) + \frac{m v^{2}}{2 k_{B}T_{b}} + \frac{\phi}{k_{B}T_{b}} \right].
\end{equation}
To avoid misunderstandings, it is worth emphasizing that no  physical meaning is given to this quantity {\em a priori}, but this issue will be considered once its dynamical behaviour is established. It can be proven  (see  \ref{ap1}) that for any solution of the BL equation it is 
\begin{equation}
\label{2.15}
\frac{\partial {\cal H(}t)}{\partial t} \leq 0,
\end{equation}
for all times. The equality only holds if
\begin{equation}
\label{2.16}
f ({\bm r},{\bm v},t)= n({\bm r},t) \varphi(v) ,
\end{equation}
where
\begin{equation}
\label{2.17}
\varphi (v)= \left( \frac{m}{2 \pi k_{B} T_{b}} \right)^{d/2} e^{-mv^{2} /2 k_{B}T_{b}}
\end{equation}
and  $n({\bm r},t)$ is an arbitrary  intruder density field. Moreover, if the two physical conditions
\begin{equation}
\label{2.18}
\int d{\bm v}\,  f({\bm r},{\bm v},t) <\infty,
\end{equation}
\begin{equation}
\label{2.19}
\int d{\bm v}\, v^{2}  f({\bm r},{\bm v},t)<\infty
\end{equation}
are verified,  and $\phi({\bm r},t)$ is bounded from below,  ${\cal H}(t)$ is also bounded from below \cite{RydL77}, implying that for any solution of the BL equation ${\cal H}(t)$ tends to a steady value ${\cal H}_{st}$. As a consequence, the probability density also tends to a stationary form  $f_{st}$. Requiring stationarity to the solution of the BL equation implies that the number density of the intruder be stationary and it has  the form
\begin{equation}
\label{2.20}
 n({\bm r})=  ce^{- \frac{\phi({\bm r})}{k_{B}T_{b}}},
 \end{equation}
with
\begin{equation}
\label{2.21}
c^{-1}= \int d{\bm r} e^{- \frac{\phi({\bm r})}{k_{B}T_{b}}} \, .
\end{equation}
Therefore, the stationary distribution, which is always reached in the long time limit, is given by the expected expression
\begin{equation}
\label{2.22}
f_{st}({\bm r},{\bm v})= n({\bm r}) \varphi(v).
\end{equation}
A short sketch of the derivation of the above property is provided in \ref{ap1}. In the steady state, it seems appropriate to identify the temperature of the intruder, assumed homogeneous, with that of the gas bath $T_{b}$.  Moreover, the steady value of the  functional ${\cal H}$ is 
\begin{equation}
\label{2.23}
{\cal H}_{st}= \ln c+ \frac{d}{2} \ln \frac{m}{2\pi k_{B}T_{b}}\, ,
\end{equation}
and it is easily seen that it accomplishes the relation
\begin{equation}
\label{2.24}
{\cal H}_{st} =- \ln Z,
\end{equation}
where $Z$ is the partition function of the intruder,
\begin{equation}
\label{2.25}
Z\equiv \int d{\bm r} \int d{\bm v}\,  e^{- \beta (\frac{mv^{2}}{2}+ \phi)} ,
\end{equation}
with $\beta \equiv (k_{B}T_{b})^{-1}$. The above results strongly suggest to identify the equilibrium free energy of the intruder as
\begin{equation}
\label{2.26}
F_{st} \equiv -k_{B} T_{b} \ln Z.
\end{equation}
The identification of $T_{b}$ as the temperature of the intruder, as well as the above
definition for the free energy  are not trivial extensions of equilibrium thermodynamics of homogenous systems to systems submitted to an external field, and they have been extensively analyzed in the literature from the perspective of ensemble theory, since they are crucial starting points for the development of the density functional theory  for inhomogeneous fluids \cite{Ev92}. A simple calculation shows that the stationary average total energy of the intruder $e_{0,st}$ can be expressed as
\begin{equation}
\label{2.27}
e_{0, st}= - \left( \frac{\partial \ln Z}{\partial \beta} \right)_{\phi} \, .
\end{equation}
 From the expression of $\ln Z$ it follows that for a quasistatic process,
\begin{equation}
\label{2.28}
dF= -k_{B}(\ln Z+\beta e_{0,eq}) dT+ \int d{\bm r}\,  n({\bm r}) \delta \phi (r),
\end{equation}
where $\delta \phi$ is the variation of the external potential, for instance,  as a consequence of the variation of an external parameter. Therefore, if one wants Eq. (\ref{1.2}) to hold as the formulation of the Second Principle for systems submitted to a nonuniform external field, we have to identify the entropy and the work as
\begin{equation}
\label{2.29}
S= k_{B}(\ln Z + \beta e_{0,eq})
\end{equation}
and
\begin{equation}
\label{2.30}
\dbar W=- \int d{\bm r} n({\bm r})\,  \delta \phi({\bm r}),
\end{equation}
respectively. Note that this definition of work is consistent with the expression used in the JR, aside from the different criteria used for the sign.  Actually, not realizing the different expressions of both $dF$ and $\dbar W$ in Eqs. (\ref{1.2}) and (\ref{2.28}) is at the origin of some discussions  about the validity of the JR appearing in the literature \cite{VyR08,HyJ08,Pe08}.  We believe that the above discussion provides a physical justification, and interpretation,  for the definition of work used in the formulation of the Jarzynski relation.

\section{Work fluctuation relations  from the Boltzmann-Lorentz equation}
\label{s3}

It is convenient to express the BL  equation in the compact form
\begin{equation}
\label{3.1}
\frac{\partial f({\bm r},{\bm v},t)}{\partial t}= \Lambda ({\bm r},{\bm v},t) f({\bm r},{\bm v},t),
\end{equation}
with 
\begin{equation}
\label{3.2}
\Lambda ({\bm r},{\bm v},t) g({\bm r},{\bm v}) \equiv -{\bm v} \cdot \frac{\partial g}{\partial {\bm r}}- \frac{\bm F}{m} \cdot \frac{\partial g}{\partial \bm v}+ J_{BL}[g,f_{b}],
\end{equation}
for arbitrary $g({\bm r},{\bm v})$. The BL equation is an evolution equation for the distribution function of the intruder. To go a little deeper into the meaning of the kinetic theory description, let us consider the mechanical Hamiltonian analysis of both the  bath particles and the intruder, assuming that the system as a whole is isolated, so all the particles obey deterministic evolution equations. Consistently with the hypothesis that the surrounding gas acts on the intruder as a thermal bath, let us assume that the initial joint probability distribution for the bath particles and the intruder factorizes in the form
\begin{equation}
\label{3.3}
\rho (\Gamma,0) =  \rho_{b}(\Gamma_{b}) f({\bm x}_{0},0),
\end{equation}
where ${\bm x} \equiv\{ {\bm r},{\bm v} \}$ denotes the phase space coordinates of the particle and $\Gamma_{b}$  is a point in the phase space associated to all the bath particles. The probability function $f({\bm x},t)$ is defined as
\begin{equation}
\label{3.4}
f({\bm x},t) \equiv  \int d \Gamma_{b} \int d{\bm x}_{0}\,  \delta \left[ {\bm x}-{\bm x}(t) \right] \rho_{b}(\Gamma_{b}) f({\bm x}_{0},0).
\end{equation}
In this expression, ${\bm x}(t)$ is the phase space point describing the dynamical state of the intruder at time $t$, assuming that at $t=0$ the point was ${\bm x}_{0}$. Of course, ${\bm x}(t)$ is determined by the deterministic equations of motion of all the particles composing the system. The form of the BL kinetic equation can be formally expressed by saying that inside phase space integrals  averaging over the initial conditions, for times large enough it is
\begin{equation}
\label{3.5}
\frac{\partial}{\partial t} \delta \left[ {\bm x}-{\bm x}(t) \right] = \Lambda( {\bm x}, t) \delta \left[{\bm x}- {\bm x}(t) \right].
\end{equation}
 Of course, this implies in particular that $f({\bm x},t)$, as defined in Eq. (\ref{3.4}), is accurately described by the BL equation. Next, define the function \cite{HyS01}
\begin{equation}
\label{3.6}
I ({\bm x},t) \equiv \int d\Gamma_{b} \int d {\bm x}_{0}\, \rho_{b}(\Gamma_{b}) f_{st}({\bm x}_{0},0) \delta \left[ {\bm x} - {\bm x}(t) \right] e^{ - \beta w^{\prime} (t)}\, ,
 \end{equation}
with the work $w^{\prime}(t)$ being given by Eq. (\ref{1.9}), and therefore it is a function of both the coordinates of the bath particles $\Gamma_{b}$ and of the intruder  ${\bm x}_{0}$.  It is 
\begin{equation}
\label{3.7}
I({\bm x},0) = f_{st}({\bm x},0). 
\end{equation}
Here and in the following we use the notation
\begin{equation}
\label{3.8}
f_{st}({\bm x},t) = Z(t)^{-1} e^{-\beta \left[ \frac{mv^{2}}{2} + \phi ({\bm r},t) \right]}\, ,
\end{equation}
\begin{equation}
\label{3.9}
Z(t) = c(t) \left( \frac{m}{2\pi k_{B}T_{b}} \right)^{-3/2},
\end{equation}
\begin{equation}
\label{3.10}
c(t)= \int d{\bm r} e^{-\beta \phi({\bm r},t)}.
\end{equation}
Time derivative of the expression of $I$  yields
\begin{equation}
\label{3.11}
\frac{\partial I}{\partial t}= -\beta \phi_{t}({\bm x},t) I+ \Lambda ({\bm x},t) I,
\end{equation}
where Eq.\ (\ref{3.5}) has been employed. Taking into account that $f_{st}({\bm x},t)$  verifies $\Lambda ({\bm x},t) f_{st}({\bm x},t)=0$, it is easily verified that the solution of the differential equation (\ref{3.11}) with the initial condition (\ref{3.7}) is
\begin{equation}
\label{3.12}
I({\bm x},t) = Z(0)^{-1} e^{-\beta \left[ \frac{mv^{2}}{2}+ \phi({\bm r},t) \right]}
\end{equation}
Integration of this expression over ${\bm x}$, taking into account the definition of $I$ given in Eq. (\ref{3.6}), gives 
\begin{equation}
\label{3.13}
\int d\Gamma_{b} \int d {\bm x}_{0} \rho_{b}(\Gamma_{b})_{b} f_{st}({\bm x}_{0} ,0) e^{- \beta w^{\prime}(t)} = \frac{Z(t)}{Z(0)}.
\end{equation}
Finally, by employing the definition of the  free energy, Eq. (\ref{2.26}), the Jarzynski relation (\ref{1.8}) follows directly.

Next, the BK relation, Eq. (\ref{1.6}), will be derived. To do so, the function
\begin{equation}
\label{3.14}
L({\bm x},t) \equiv  \int d\Gamma_{b} \int d {\bm x}_{0}\, \rho_{b}(\Gamma_{b}) f_{st}({\bm x}_{0},0) \delta \left[ {\bm x} - {\bm x}(t) \right] e^{ - \beta w (t)}\, ,
 \end{equation}
is introduced. The work $w(t)$ is defined by Eq. (\ref{1.7}), i.e.
\begin{equation}
\label{3.15}
w(t) = -\int_{0}^{t} d\tau\, {\bm v}(\tau) \cdot \phi_{1 {\bm r}}\left[ {\bm x}(\tau),\tau \right],
\end{equation}
with
\begin{equation}
\phi_{1 {\bm r}}\left[ {\bm x},\tau \right] \equiv \left( \frac{\partial \phi_{1} ({\bm r},\tau)}{\partial {\bm r}} \right)_{\tau}.
\end{equation} 
From Eq. (\ref{3.14}) it follows that
\begin{equation}
\label{3.16}
L({\bm x},0)= f_{st}({\bm x},0).
\end{equation}
Consider
\begin{equation}
\label{3.17}
\int_{0}^{t} d\tau \frac{d}{d \tau}  \phi_{1}\left[ {\bm x}(\tau),\tau \right] = \int_{0}^{t} d\tau\,  \left\{ \phi_{\tau} \left[ {\bm x}(\tau),\tau\right] +{\bm v}(\tau) \cdot \phi_{1 {\bm r}}\left[ {\bm x}(\tau),\tau \right] \right\},
\end{equation}
and, since $\phi_{1}({\bm x},0)=0$,
\begin{equation}
\label{3.18}
\phi_{1}[{\bm x}(t),t] = w^{\prime}(t)- w(t).
\end{equation}
Therefore, Eqs.\, (\ref{3.6}) and (\ref{3.14}) give
\begin{equation}
\label{3.19}
L({\bm x},t)= e^{\beta \phi_{1}({\bm x},t)} I({\bm x},t)= \frac{e^{- \beta \left[ \frac{mv^{2}}{2}+ \phi_{0}({\bm x}) \right] }}{Z(0)}.
\end{equation}
In the last transformation, Eq. (\ref{3.12}) has been used. Integration of the above equality with respect to ${\bm x}$ leads to the desired result,
\begin{equation}
\label{3.20}
\int d\Gamma_{b} \int d{\bm x}_{0}\,  \rho_{b}(\Gamma_{b}) f_{st}({\bm x}_{0},0) e^{-\beta w(t)}=1.
\end{equation}
Let us emphasize that Eq.\ (\ref{3.19}) shows that both work fluctuation relations, although apparently very different, are closely related.  Also, it is worth stressing that the functions $I$ and $L$ remain Maxwellian, with the $\beta$ parameter determined by the bath temperature, for all times and then the collision term in Eq. (\ref{3.11}) vanishes.

\section{Numerical simulations of the kinetic equation}
\label{s4}
In order to investigate whether the above theoretical predictions are easy to observe, in the sense of how many trajectories are needed to get reliable results, and also to study the work probability distributions for both definitions (Jarzynski and Bochkov and Kuzovlev), the kinetic equation has been solved using the direct simulation Monte Carlo  (DSMC) method 
 \cite{Bi94}. This is a particle simulation method, in which the actual dynamics of the particles is substituted by an effective stochastic dynamics consistent with the low density  limit. It has been rigorously proven that the average over trajectories  provides a solution of the Boltzmann equation. The method, originally designed for the nonlinear Boltzmann equation, can be easily adapted for the BL equation \cite{BRGyD99}. In the simulations to be reported, hard-sphere interactions of diameter $d$ between the intruder and the gas particles have been employed.  Moreover, the mass of the intruder has been taken the same as that of the bath particles, i.e. $m=m_{b}$. Two different external fields have been employed. In case I, an harmonic potential is perturbed by a uniform force whose amplitude grows linearly in time. More specifically,
\begin{equation}
\label{4.1}
\phi_{0}({\bm x})= \frac{m \omega_{0}^{2} x^{2}}{2}
\end{equation}
and
\begin{equation}
\label{4.2}
\phi_{1}({\bm x},t)= -f_{0}\frac{t}{t_{0}} x \Theta(t).
\end{equation}
In case II, the unperturbed potential $\phi_{0}({\bm x})$ is the same as in case I, and $\phi_{1}$ is another harmonic field,
\begin{equation}
\label{4.3}
\phi_{1}({\bm x},t)=  \frac{m \omega_{1}(t)^{2} x^{2}}{2}
\end{equation}
with
\begin{equation}
\label{4.4}
\omega_{1}^{2}(t)= \omega_{1f}^{2} \frac{t}{t_{0}}\, \Theta (t)
\end{equation}
 In the above expressions, $w_{0}$, $f_{0}$, $t_{0}$, and $w_{1f}$ are constants to be specified later, and $\Theta (t)$ is the Heaviside step function. The time parameter $t_{0}$ controls how fast the perturbation is applied, the limit $t_{0} \rightarrow \infty$ defining the quasistatic process. Notice that all the forces act  along the same direction, namely along the $x$ axis. 
 
 The simplicity of the chosen external fields allows to evaluate analytically the partition function defined in Eq. (\ref{2.25}) and hence to get the value of the equilibrium free energy  associated to each value of $\phi({\bm x},t)$ by means of Eq. (\ref{2.26}).  In the simulations, the time origin is always taken after the system has reached a stationary state with the harmonic potential $\phi_{0}$. The form for the external potentials was motivated by comparison purposes, since these potentials have been used previously in the literature \cite{VyR08,HyJ08}.  The reported results have been averaged over $10^{7}$ trajectories, and dimensionless quantities have been defined by taking  the mean free path of the gas particles, $\lambda$, as unit of length,  the mass of the gas particles $m$, as the unit of mass, and $k_{B}T_{b}$ as the energy unit. 
 
\begin{figure} 
\begin{center}
\includegraphics[width=.6\textwidth]{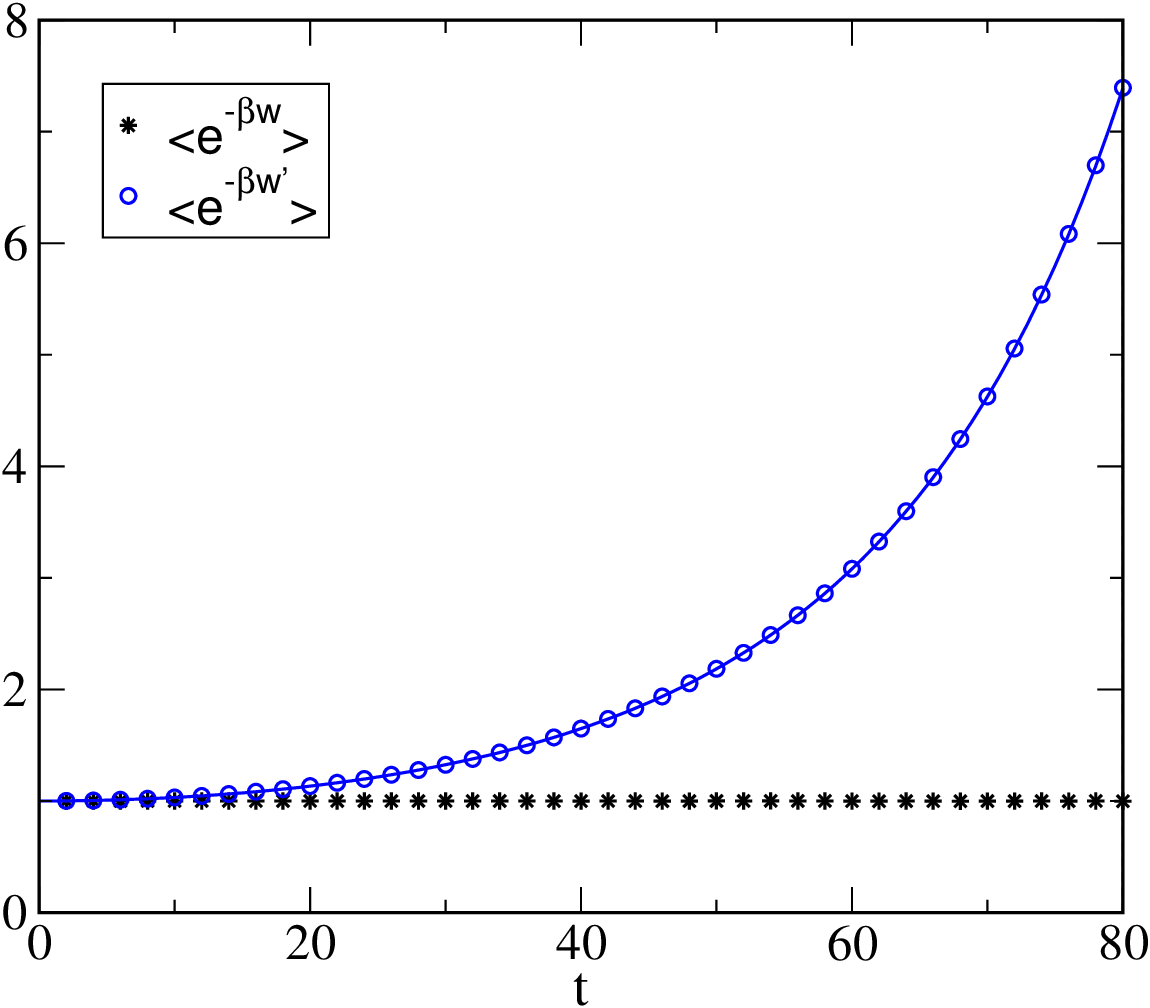}
 \caption{Time evolution obtained with the  DSMC method of the Bochkov and Kuzovlev work function, Eq. (\ref{1.6}),  (stars) and the Jarzynski work function, Eq. (\ref{1.8}) 
 (circles), for an intruder immersed in a low density gas at equilibrium described by the Boltzmann-Lorentz kinetic equation. All particles are hard spheres. Time $t$ is measured in the dimensionless units defined in the main text. The constant potential and the time-dependent  perturbation are given by Eqs. (\ref{4.1}) and (\ref{4.2}), respectively. The solid line is the exact theoretical value for the JR.} \label{fig1} \end{center}
 \end{figure}

\begin{figure}  
\begin{center}
\includegraphics[width=.6\textwidth]{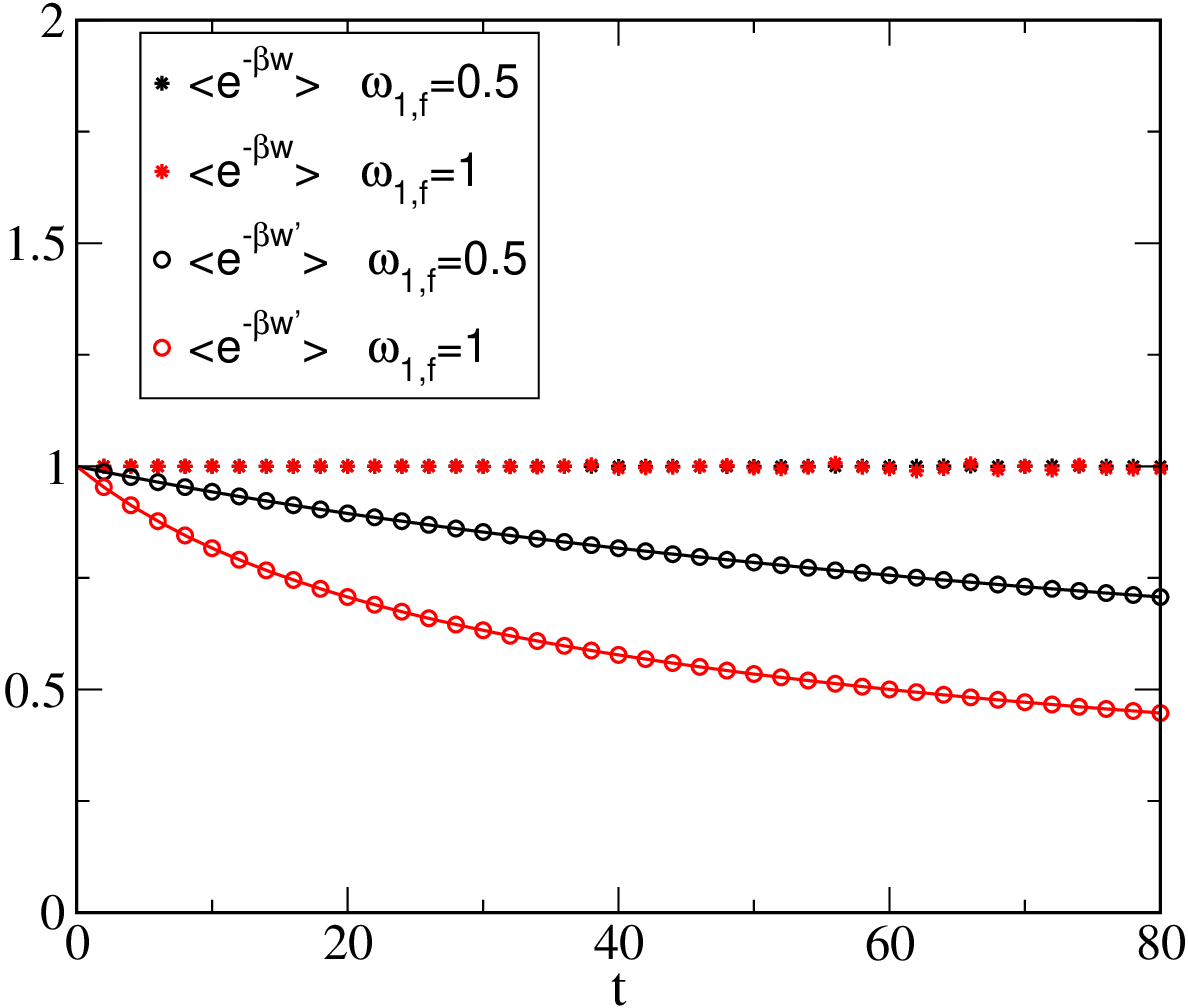} 
\caption {The same as in Fig.\ \protect{\ref{fig1}} but now for the potentials given by Eqs.\ (\ref{4.1}) and (\ref{4.3}). The (red and black) stars for the Bochkov and Kuzovlev DSMC
results corresponding to the two values of $\omega_{1f}$ coincide over the scale of the figure.}  \label{fig2}
\end{center}
\end{figure}

 In Fig.\ \ref{fig1}, the average values of $e^{-\beta  w (t)}$  and of $e^{-\beta w^{\prime} (t)}$ are plotted as  functions of time for the perturbation referred to as  case I. The values of the parameters are  $\omega_{0}=0.5$, $f_{0}=1$, and $ t_{0}=80$. Symbols are simulation results, while the solid line is the theoretical prediction of the JR, using the values of the free energy obtained analytically from Eqs. (\ref{2.25}) and (\ref{2.26}). It is observed that both work theorems are quite well fulfilled by the numerical data. A similar conclusion is reached for the perturbation corresponding to case II as it can be observed in the results shown in Fig. \ref{fig2}. In the reported results, two different values of the final  frequency of the perturbation, $w_{1f}$, have been employed, as indicated in the inset of the figure. 
 
Consider the Jarzynski definition of work, and define  the probability density, $P(w^{\prime},t)$, of getting a given value for it along a given protocol of variation of the external field, so that
 \begin{equation}
 \label{4.5}
 \langle e^{-w^{\prime}(t)/k_{B}T_{b}} \rangle = \int dw^{\prime}\, P(w^{\prime},t) e^{-w^{\prime}/k_{B}T_{b}},
 \end{equation}
 and similarly for any other function of $w^{\prime}(t)$. Let us introduce the joint probability density, $P({\bm x},w^{\prime},t)$,  for given values of the position and velocity of the intruder at time $t$, and the work carried out up to that time, along a given protocol of variation of the external potential. This quantity is given by
 \begin{equation}
 \label{4.6}
 P({\bm x},w^{\prime},t) = \int d \Gamma_{b} \int d{\bm x}_{0}\, \rho_{b}(\Gamma_{b}) f_{st}({\bm x}_{0},0) \delta \left[{\bm x}-{\bm x}(t) \right] \delta \left[ w^{\prime}- w^{\prime}(t) \right],
 \end{equation}
 where once again it has been assumed that the intruder was at equilibrium at $t=0$, when the perturbation is switched on. Trivially it is
 \begin{equation}
 \label{4.7}
\int d{\bm x}\, P({\bm x},w^{\prime},t)= P(w^{\prime},t).
 \end{equation}
 From Eqs. (\ref{4.6}) and  (\ref{3.5}) it follows that 
 \begin{equation}
 \label{4.8}
 \frac{\partial}{\partial t}\, P({\bm x},w^{\prime},t) = \Lambda ({\bm x},t) P({\bm x},w^{\prime},t) +\phi_{t}({\bm x},t) \frac{\partial}{\partial w^{\prime}}\, P({\bm x},w^{\prime},t).
 \end{equation}
 This differential equation is to be solved with the initial condition 
 \begin{equation}
 \label{4.9}
 P({\bm x},w^{\prime},0)= f_{st}({\bm x},0) \delta (w^{\prime}).
 \end{equation}
 An analogous equation can be derived for the joint distribution of ${\bm x}$ and the work along a trajectory $w(t)$ considered by Bochkov and Kuzovlev. Nevertheless, both equations are hard to solve for nontrivial external potentials, so in the following  numerical results obtained by the DSMC method will be reported. 
 
\begin{figure}
\begin{center}
\includegraphics[width=.6\textwidth]{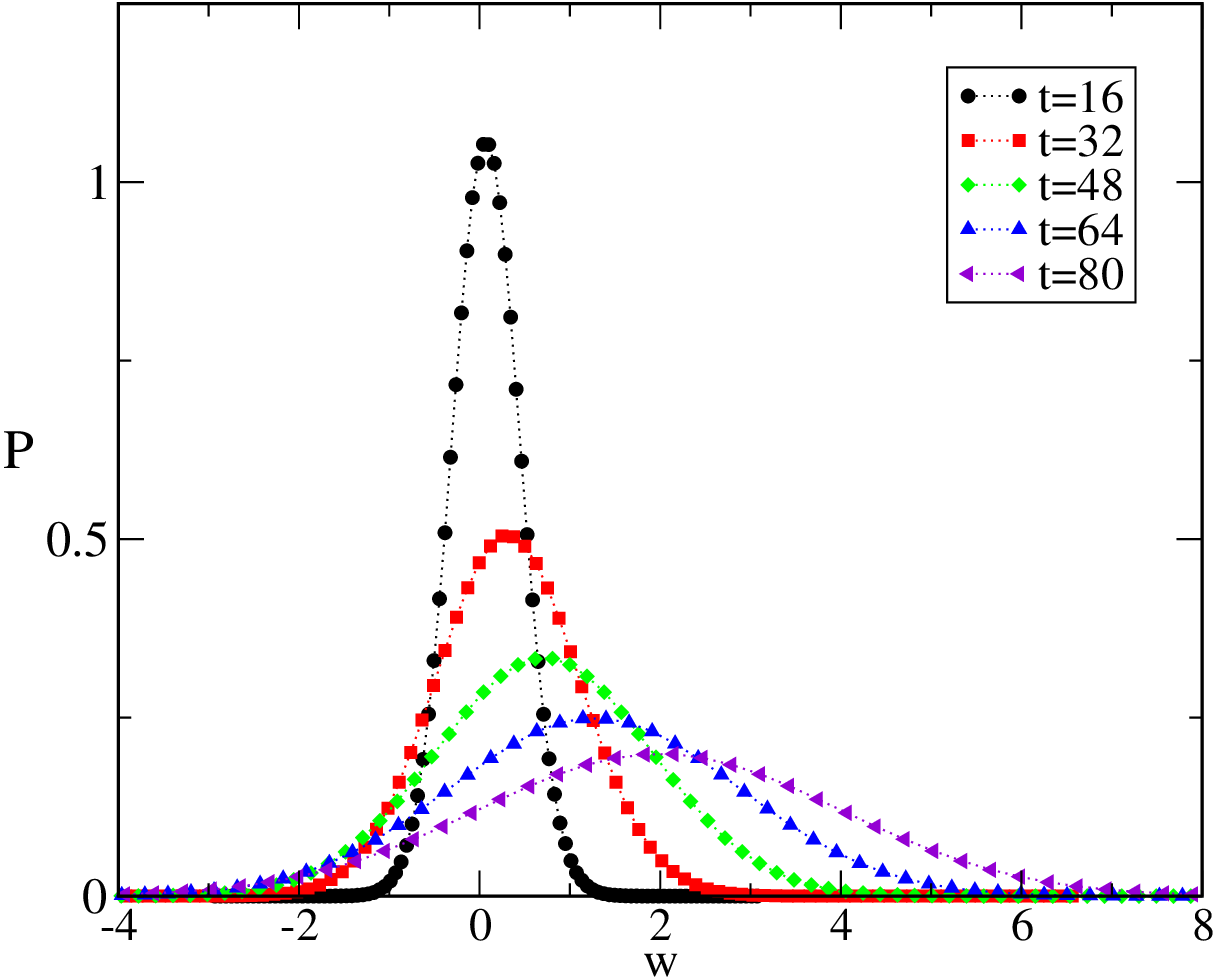}
  \caption{ Bochkov and Kuzovlev work  distribution for the same system as considered in Fig.\ \protect{\ref{fig1}}. The different symbols correspond to DSMC results at five different times, as indicated in the inset.  As time increases the curves move to the right.}
  \label{fig3}\end{center}
\end{figure}

\begin{figure}\begin{center}
\includegraphics[width=.6\textwidth]{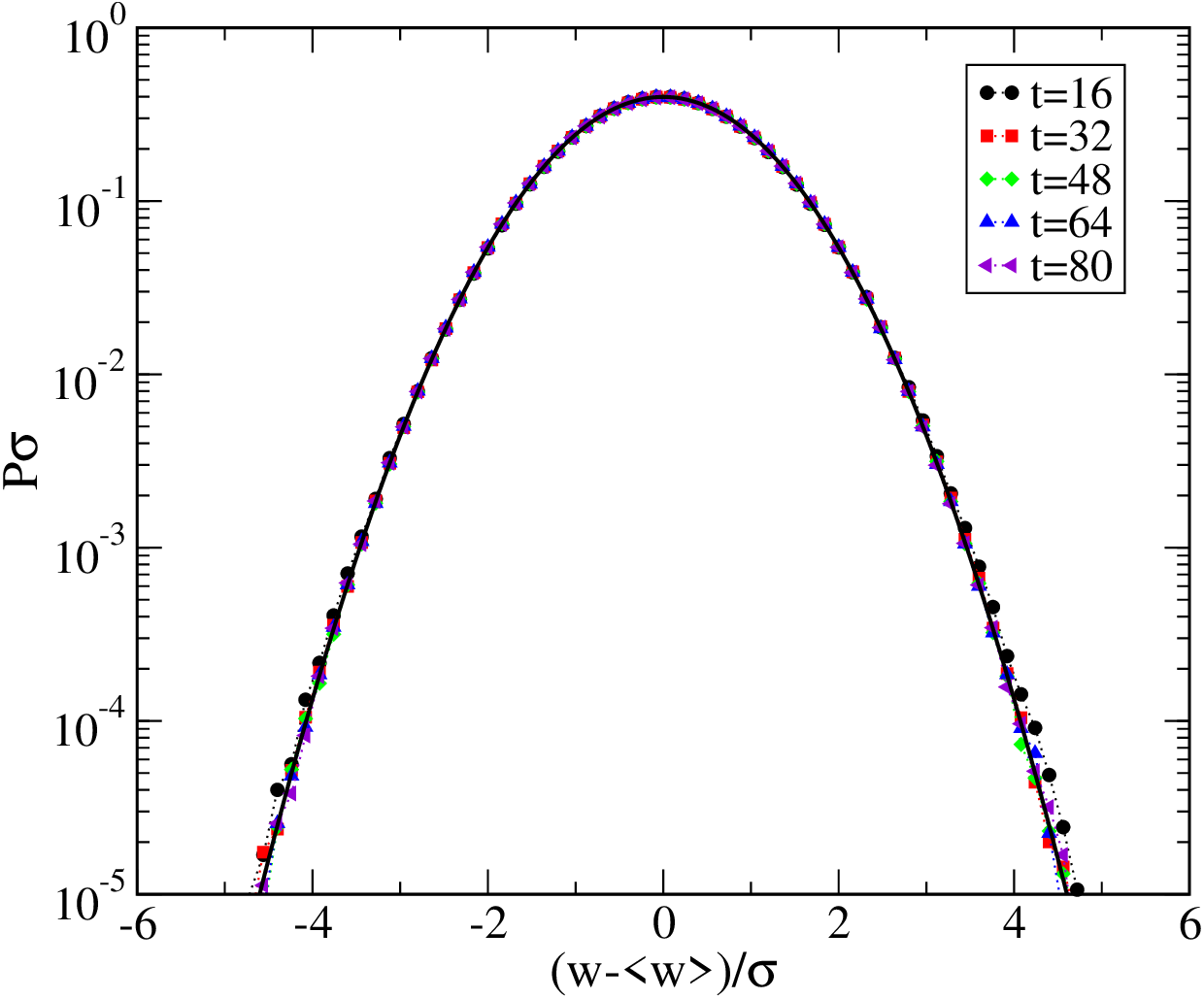} 
\caption { The same as in Fig. \protect{\ref{fig3}} but now each of the curves is scaled with its standard deviation and displaced its mean value. Moreover a logarithmic representation is employed. The solid line is the Gaussian distribution with zero mean and unit standard deviation.}\end{center}
  \label{fig4}
\end{figure}

 In Fig.\ \ref{fig3}, the time evolution of the  probability distribution of the BK expression of work $w$  at different times is shown for the same system as in Fig. \ref{fig1}. It is observed that as time progresses the width of the distribution increases and its maximum moves to the right, i.e.  positive  values of the work become more frequent. Actually, the distribution seems to be  Gaussian at all times, as seen in Fig.\, \ref{fig4},  where the distributions of $ \left( w-\langle w \rangle  \right)/ \sigma$, with $\sigma$ being the standard deviation  of each original distribution, are plotted on a logarithmic scale. 
 
\begin{figure}\begin{center}
\includegraphics[width=.6\textwidth]{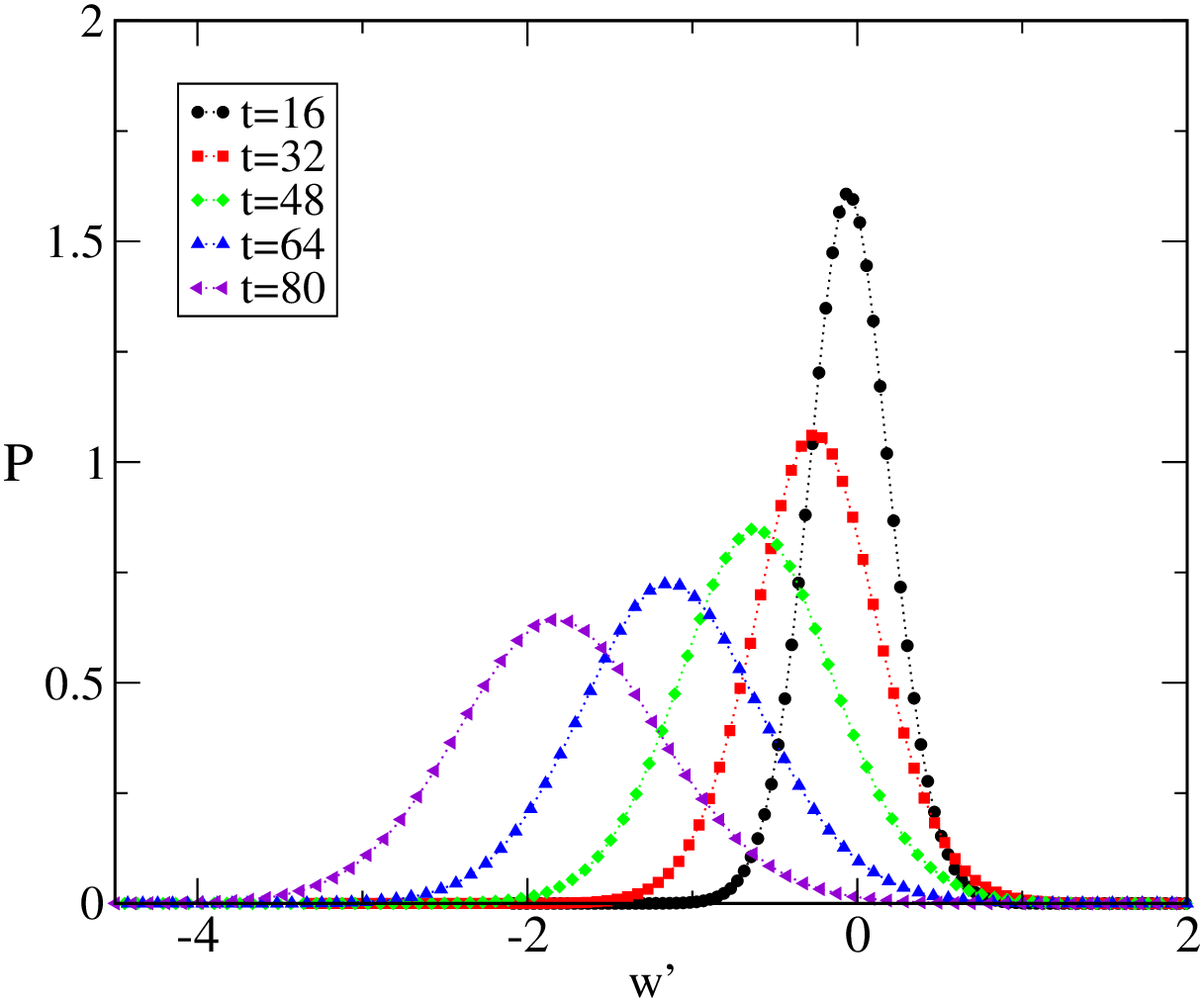}
  \caption{Jarzynski work  distribution for the same system as considered in Fig.\ \protect{\ref{fig1}}. The different symbols correspond to DSMC results at five different times, as indicated in the inset.  As time increases the curves move to the left.}
  \label{fig5}\end{center}
\end{figure}

\begin{figure}\begin{center}
\includegraphics[width=.6\textwidth]{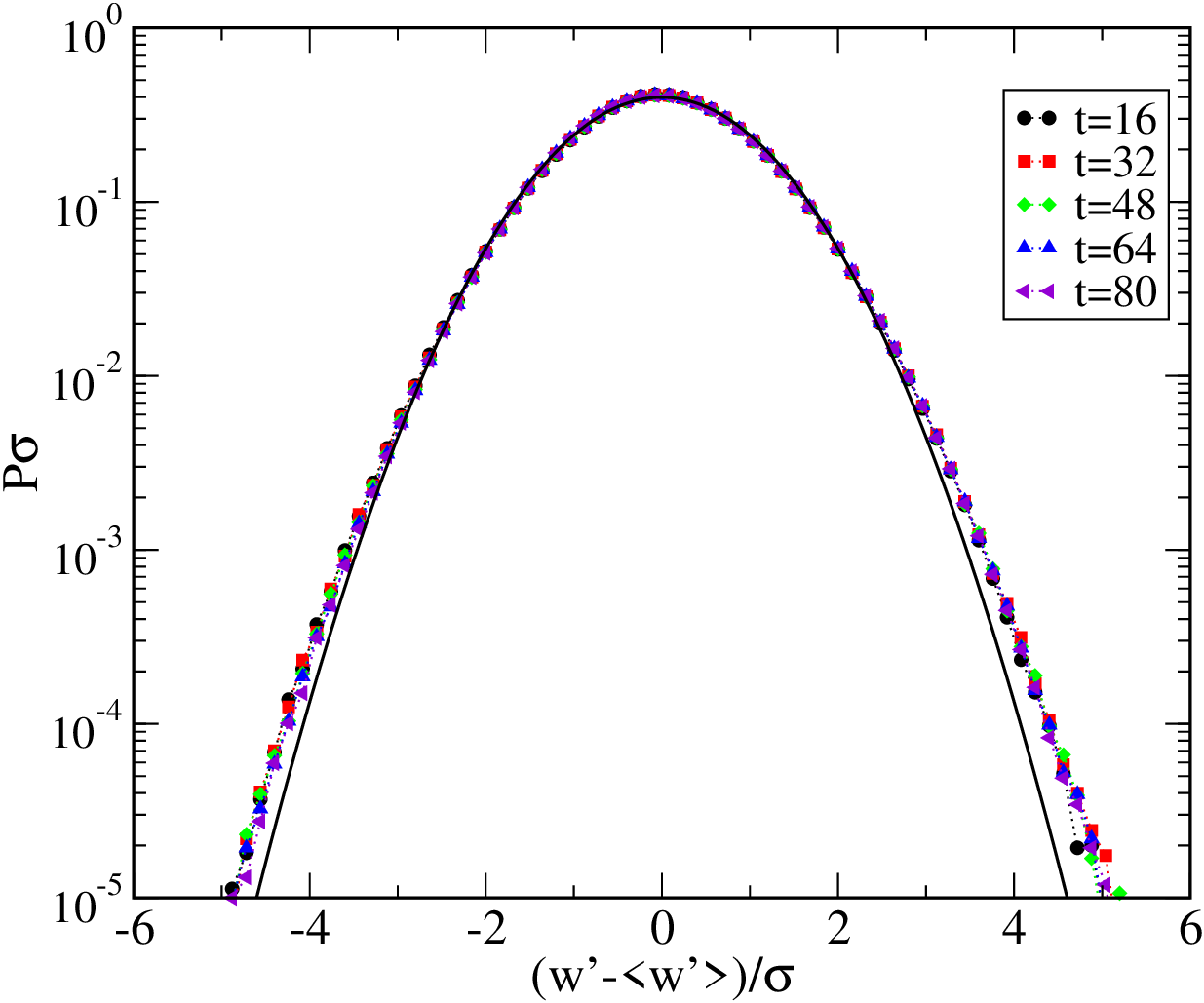} 
\caption { The same as in Fig.\protect{\ref{fig5}} but now each of the curves is scaled with its standard deviation and displazed its mean value. Moreover a logarithmic representation is employed. The solid line is the Gaussian distribution with zero mean and unit standard deviation.}
  \label{fig6} \end{center}
\end{figure}

In the case of the work definition used by Jarzynski, the behaviour of the probability distribution is similar, but with two key differences, as it can be observed in Figs.\ \ref{fig5} and \ref{fig6}. First, as time increases the curves move to the left, i.e. negative values of the work are more frequent. The second difference is that now the distributions seem to be clearly non-Gaussian since the deviation observed at both tails of the distribution in Fig. \ref{fig6} can hardly be attributed to statistical uncertainties, given the systematic character of the deviations. In any case, the sharp collapse of the curves when scaling must be noticed.
 
\begin{figure}\begin{center}
\includegraphics[width=.6\textwidth]{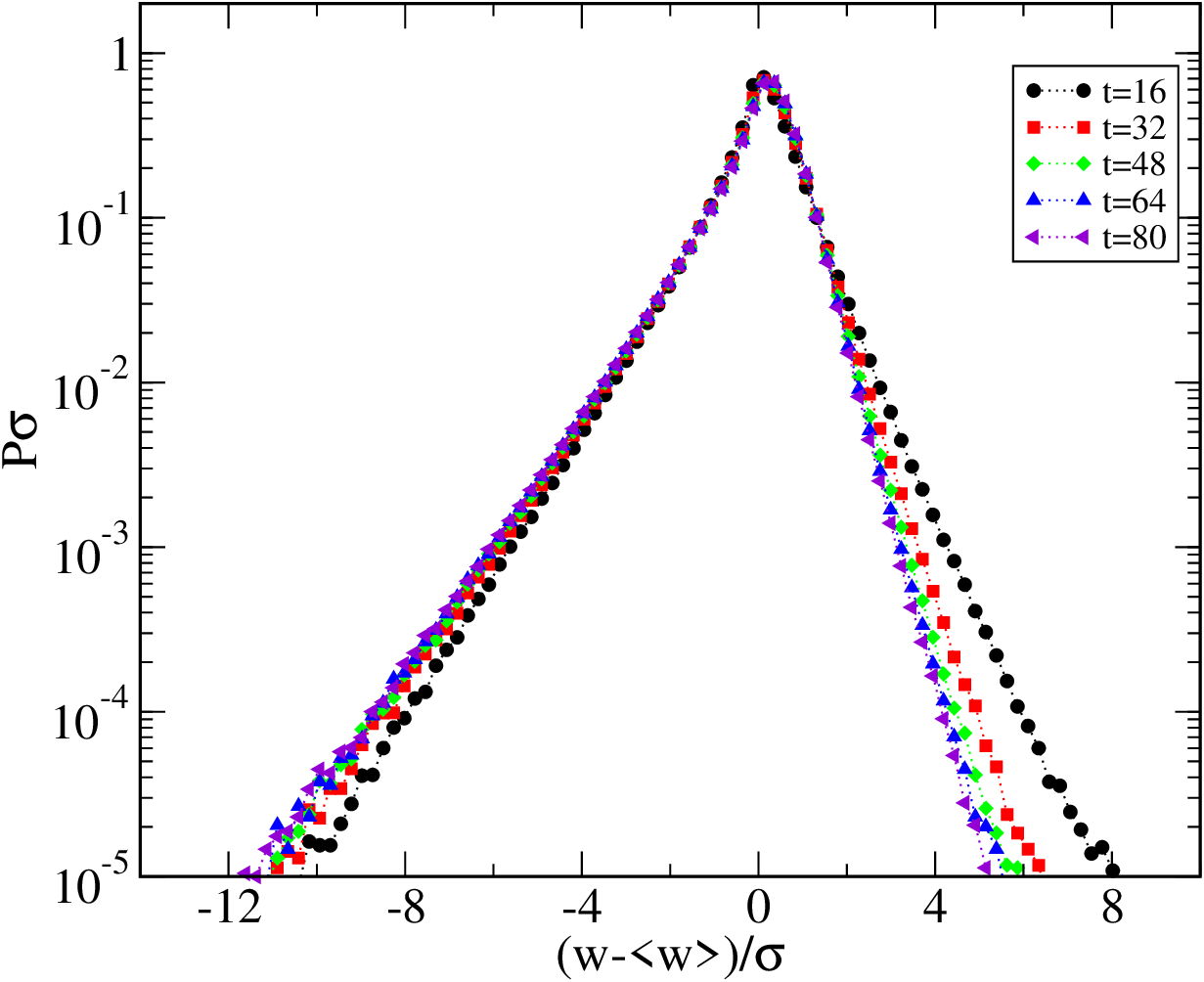}
  \caption{Bochkov and Kuzovlev work distribution for the same system as considered in Fig. \ref{fig2}. The different symbols correspond to DSMC results at five different times, as indicated in the inset. Each of the curves is scaled with its standard deviation and displaced an amount equal to the work mean value. Moreover a logarithmic representation is employed. }
  \label{fig7}\end{center}
\end{figure}

\begin{figure}\begin{center}
\includegraphics[width=.6\textwidth]{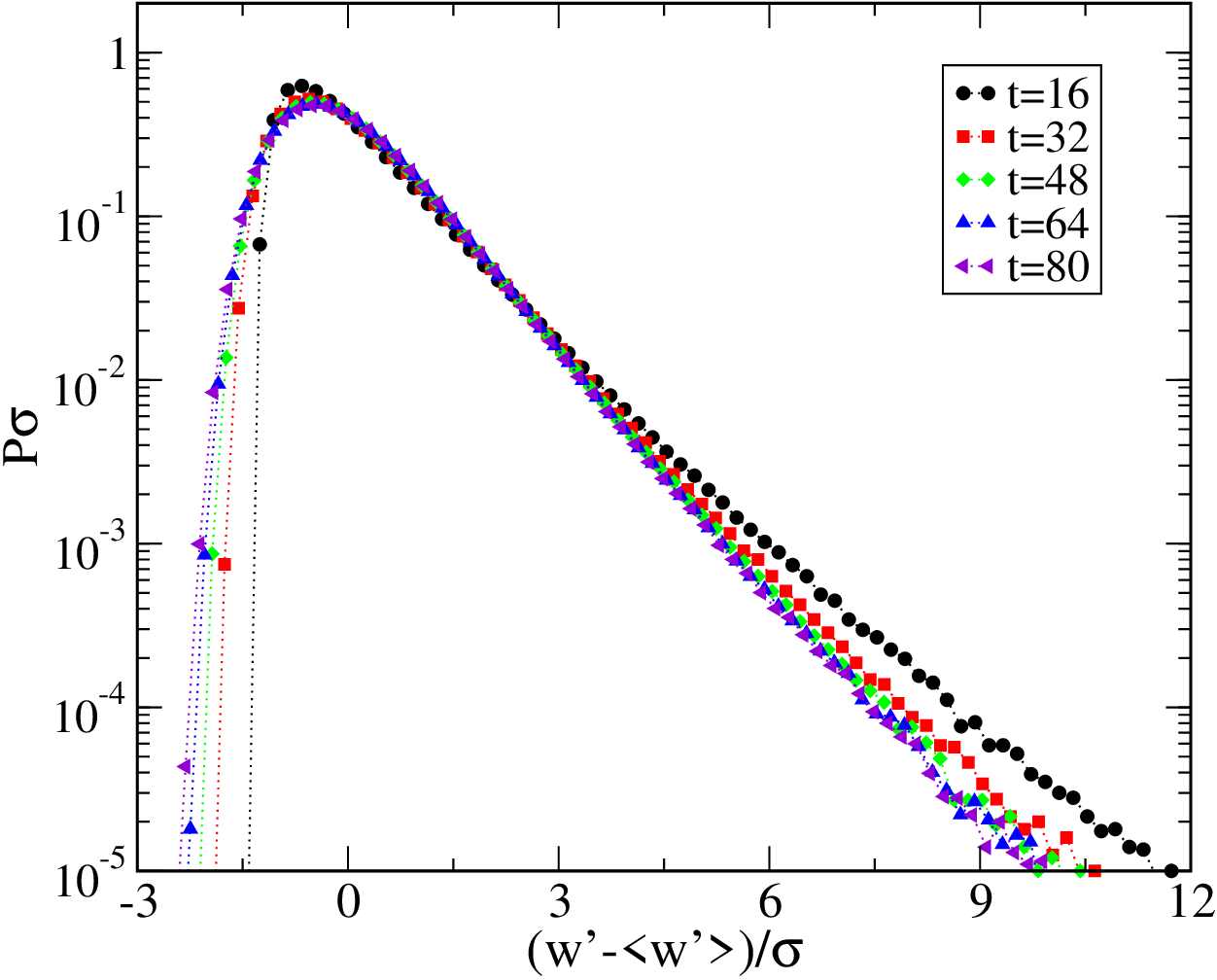} 
\caption{ Jarzynski  work distribution for the same system as considered in Fig. \ref{fig2}. The different symbols correspond to DSMC results at five different times, as indicated in the inset. Each of the curves is scaled with its standard deviation and displaced an amount equal to the work mean value. Moreover a logarithmic representation is employed.}
  \label{fig8}\end{center}
\end{figure}

We have performed the same study for case II, i.e. for external potentials given by Eqs.\ (\ref{4.1}) and (\ref{4.3}). The results reported in Figs. \ref{fig7} and \ref{fig8} are for a system with the same values of the parameters as in Fig.\,  \ref{fig2}, but only the value $\omega_{1f}=1$ is displayed. It follows from the figures that the scaling does not collapse the curves for this perturbation. Moreover, the curves strongly deviate from a Gaussian and exhibit exponential tails.
The conclusion is that the shape of the work distributions strongly depends on the definition of work used and on the particular external perturbation applied to the system. These features were expected. Something more surprising is that the shape of the work distribution for a given external potential changes in time in a  nontrivial way, in spite of the fact that the two work fluctuation relations we are studying, which refer to  the average of exponential functions, hold for all times.

\begin{figure}\begin{center}
\includegraphics[width=.6\textwidth]{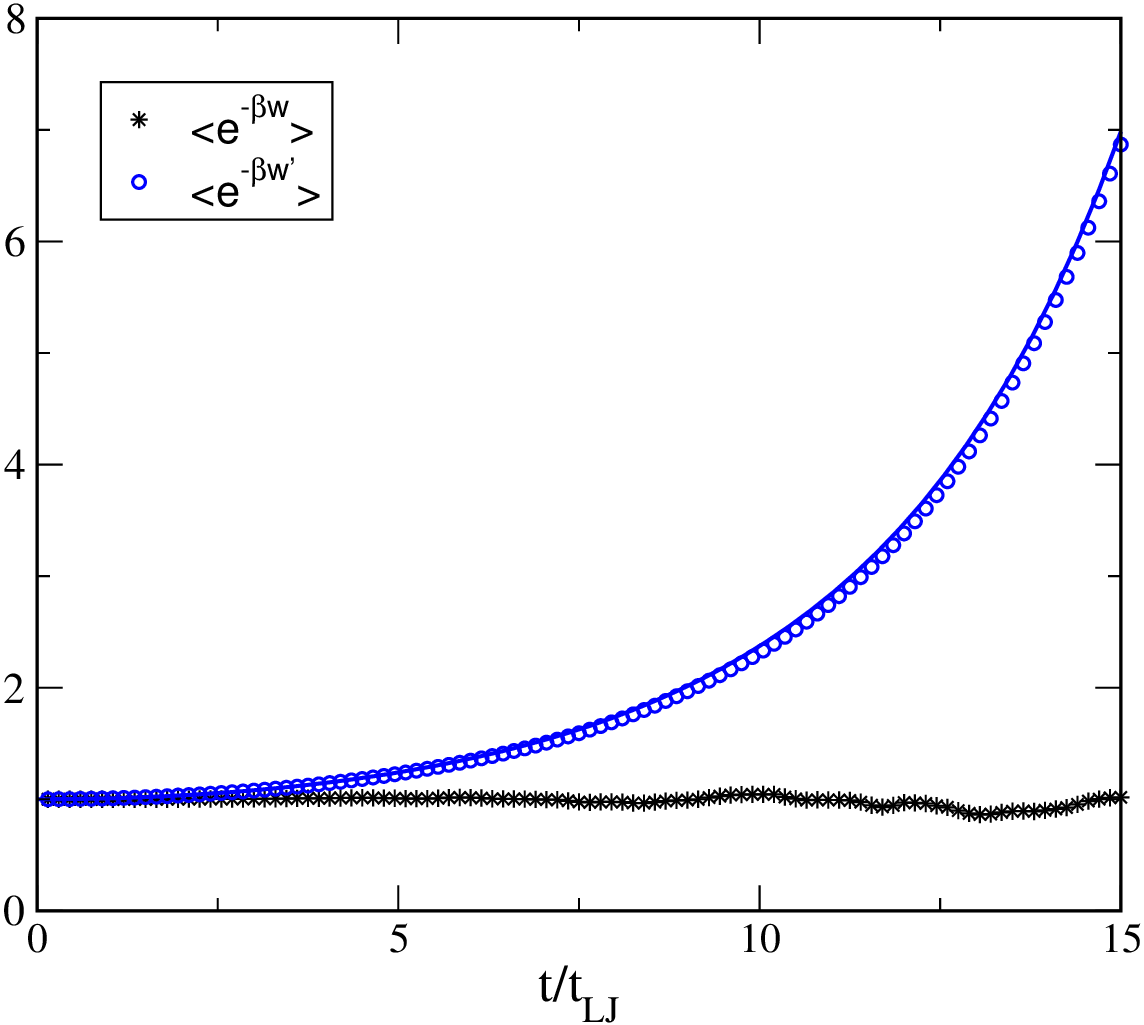} 
 \caption{ MD results for the BK function (stars) and the  Jarzynski work function (circles) in a Lennard-Jones system with $n\sigma^{3}=0.3$. The external potential 
 for the intruder was case 1, with  $f_{0}/(m\omega_{0}^{2}\sigma)=4$, $t_{0}/t_{LJ}=15$.   The solid line is the exact theoretical value for   $e^{-\beta \Delta F}$.}\label{fig9} \end{center}
\end{figure}

To test the actual accuracy of the theoretical predictions derived from the BL equation, we have also performed Molecular Dynamics (MD)
simulations of a tagged particle immersed in a bath of identical particles, so that the explicit form of the kinetic equation is not assumed. The particles interact by a Lennard-Jones potential of diameter $\sigma$ and depth of the attractive
well $\epsilon$. As in the previous simulations, only the tagged particle feels the external potentials, that were chosen identical to those of the DSMC study, cases I and II. 
In our MD simulations,
a system of $N=1000$ particles was considered, and the results were averaged over $4000$ trajectories. Three different values of the density were
investigated, namely $n\sigma^{3}=0.1$, $0.3$, and $0.5$.    In Fig. \ref{fig9}
the MD results for the averages of both  $e^{-\beta  w (t)}$  and of $e^{-\beta w^{\prime} (t)}$ are plotted as a function of $t/t_{LJ}$, with $t_{LJ}=\sigma (m/\epsilon)^{1/2}$ for the
perturbation named case I. The density in this case was $n\sigma^{3}=0.3$, and the external potential parameters were $f_{0}/(m\omega_{0}^{2}\sigma)=4$, $t_{0}/t_{LJ}=15$.   
The solid line is the exact theoretical values for $e^{-\beta \Delta F}$. As it happened with the DSMC simulations, the simulation results are in very good agreement with the
two, BK and JR, theorems. It is noticed in the MD simulations that the results for the averages are noisier than in the DSMC case, but this is because in the MD simulations results
are averaged over 4000 trajectories, while in DSMC $10^{7}$ trajectories of the tagged particle were considered. For all the cases we have  studied, the results obtained with MD
are identical to those obtained with DSMC, apart from the larger noise in the former.

\section{Summary and final comments}
\label{s5}
It has been shown that both the Bochkov and Kuzovlev relation and the Jarzynski relation, are fulfilled by a particle or intruder immersed in a much larger dilute system at equilibrium. Although the theoretical results presented here are restricted to the BL kinetic equation, we have also performed molecular dynamics (MD) simulations at low density, and obtained fully consistent results.

On the other hand, it must be emphasized that the required measurements of the fluctuations of the respective works involved in each relation seem  hard tasks in practice. The order of magnitude of the number of trajectories required to obtain a result with low noise level is very high, at least several thousands in the simulations we have performed (DSMC and MD). This difficulty has already been pointed out in the literature \cite{KKTyY12,TMYyH12,KDyS16}. Consequently, it is hard to see any advantage of this procedure over measuring the work in the quasistatic limit of an isothermal process, in order  to measure equilibrium free energy changes.

It has been shown that on the basis of kinetic theory it is possible to formulate a well founded non-equilibrium macroscopic theory for a particle in contact with a heat bath. This approach can be  a complementary alternative to the so-called stochastic thermodynamics.

The analysis presented here can be directly extended to systems described by a linear kinetic theory. This extension can be seen to be trivial for all tagged particle kinetic equations with a  collision term that vanishes for Mawellians with the appropriate temperature parameter. A nontrivial and interesting extension, surely requiring a more complex analysis,  is to consider nonlinear kinetic equations, e.g. the Boltzmann and Enskog equations.

\ack

This research was supported by the Ministerio de Econom\'{\i}a y Competitividad  (Spain) through Grant No. FIS2014-53808-P (partially financed by FEDER funds).

\appendix

\section{The {\cal H} theorem for the Boltzmann-Lorentz equation in an external field}
 \label{ap1}
In this appendix a short outline of the derivation of the theorem stated in Sec. \ref{s2} is provided.  Taking time derivative in Eq. (\ref{2.14}) yields
\begin{equation}
\label{a1.1}
\frac{\partial {\cal H}}{\partial t}= \int d{\bm r} \int d{\bm v}\,  \frac{\partial f}{\partial t} \left(  \ \ln f + \frac{m v^{2}}{2 k_{B}T_{b}} + \frac{\phi}{k_{B}T_{b}} \right).
\end{equation}
The BL kinetic equation is decomposed in the form
\begin{equation}
\label{a1.2}
\frac{\partial f}{\partial t}= \left( \frac{\partial f}{\partial t} \right)_{ flux}+J_{BL}[f,f_{b}],
\end{equation}
with
\begin{equation}
\label{a1.3}
\left( \frac{\partial f}{\partial t} \right)_{flux} \equiv - {\bm v} \cdot \frac{\partial f}{\partial {\bm r}} - \frac{\bm F}{m}\, \cdot  \frac{\partial f}{\partial {\bm v}}\, .
\end{equation}
Consider first
\begin{equation}
\label{a1.4}
\left(\frac{\partial {\cal H}}{\partial t}\right)_{flux}\equiv  \int d{\bm r} \int d{\bm v}\,  \left( \frac{\partial f}{\partial t} \right)_{flux}  \left(  \ \ln f + \frac{m v^{2}}{2 k_{B}T_{b}} + \frac{\phi}{k_{B}T_{b}} \right).
\end{equation}
A simple calculation, assuming that the system is closed and isolated in the sense that there is no flux of particles or any other property through the boundaries, and that the distribution function $f$ decays fast enough for large values of the velocity, as it is usually done, leads to
\begin{equation}
\label{a1.5}
\left(\frac{\partial {\cal H}}{\partial t}\right)_{flux} =0.
\end{equation}
Therefore,
\begin{equation}
\label{a1.6}
\frac{\partial {\cal H}}{\partial t}= \int d{\bm r} \int d{\bm v}\,  J_{BL}[{\bm r},{\bm v},t|f,f_{b} ] \left(  \ln f + \frac{m v^{2}}{2 k_{B}T_{b}} + \frac{\phi}{k_{B}T_{b}} \right).
\end{equation}
The BL collision term  verifies
\begin{eqnarray}
\label{a1.7}
\int d{\bm v} a({\bm v}) J_{BL}[{\bm r},{\bm v},t|f,f_{b} ] =& \nonumber\\
 \int d{\bm v} \int d{\bm v}_{1} \int d \Omega\, \left[ a({\bm v}^{\prime})-a({\bm v}) \right]  \sigma (\Omega,g) g f({\bm r},{\bm v},t)f_{b}(v_{1}), &
\end{eqnarray}
for any arbitrary function $a({\bm v})$. This relation follows from the properties of elastic collisions, namely the volume conservation in velocity space, the equality of the cross section for a collision and its inverse, and the conservation of the module of the relative velocity. Use of the property (\ref{a1.7}) leads to

\begin{eqnarray}
\label{a1.8} 
\int d{\bm r}  \int d{\bm v}\,  J_{BL}[{\bm r},{\bm v},t|f,f_{b} ] \left(  \frac{m v^{2}}{2 k_{B}T_{b}} + \frac{\phi}{k_{B}T_{b}} \right)  =& \nonumber \\
\fl - \int d {\bm r} \int d{\bm v} \int d{\bm v}_{1} \int d\Omega\,  \sigma (\Omega,g) g  \left[ f({\bm r},{\bm v}^{\prime},t) f_{b}(v^{\prime}_{1})-f({\bm r},{\bm v},t) f_{b}(v_{1}) \right] \frac{m_{b}v_{1}^{2}}{2k_{B}T_{b}} =  & \nonumber \\
 \fl \int d {\bm r} \int d{\bm v} \int d{\bm v}_{1} \int d\Omega\,  \sigma (\Omega,g) g  \left[ f({\bm r},{\bm v}^{\prime},t) f_{b}(v^{\prime}_{1})-f({\bm r},{\bm v},t) f_{b}(v_{1}) \right] \ln f_{b}(v_{1}),&
\end{eqnarray}
and substitution of this result into Eq. (\ref{a1.6})  gives
\begin{eqnarray}
\label{a1.9}
\frac{\partial {\cal H}}{\partial t}&= &    \int d {\bm r} \int d{\bm v} \int d{\bm v}_{1} \int d\Omega\,  \sigma (\Omega,g) g  \left[ f({\bm r},{\bm v}^{\prime},t) f_{b}(v^{\prime}_{1}) \right. \nonumber \\ 
&&  - \left. f({\bm r},{\bm v},t) f_{b}(v_{1}) \right] \ln \left[ f({\bm r},{\bm v},t) f_{b}(v_{1}) \right]   \nonumber \\ 
&=&         \frac{1}{2} \int d {\bm r} \int d{\bm v} \int d{\bm v}_{1} \int d\Omega\,  \sigma (\Omega,g) g  \left[ f({\bm r},{\bm v}^{\prime},t) f_{b}(v^{\prime}_{1}) \right. \nonumber \\
&& - \left. f({\bm r},{\bm v},t) f_{b}(v_{1}) \right] \ln \frac{f({\bm r},{\bm v},t) f_{b}(v_{1})}{f({\bm r},{\bm v}^{\prime},t) f_{b}(v_{1}^{\prime})} \leq 0.
\end{eqnarray}
 The equality sign only holds if $f({\bm r},{\bm v},t) =f_{l}({\bm r},{\bm v},t)$ such that
 \begin{equation}
 \label{a1.10}
  \frac{f_{l}({\bm r},{\bm v},t) f_{b}(v_{1})}{f_{l}({\bm r},{\bm v}^{\prime},t) f_{b}(v_{1}^{\prime})}=1,
  \end{equation}
i.e.,
\begin{equation}
\label{a1.11}
f_{l}({\bm r},{\bm v},t)= n({\bm r},t) \varphi (v),
\end{equation} 
 where $\varphi (v)$ is given by Eq. (\ref{2.17}) and $n({\bm r},t)$ is up to this point arbitrary, aside from the normalization condition. Moreover, if the two conditions (\ref{2.18}) and (\ref{2.19}) are verified, ${\cal H}(t)$ is bounded from below \cite{RydL77} and
 \begin{equation}
 \label{a1.12}
 \lim_{t \rightarrow \infty} f({\bm r},{\bm v},t) = f_{l}({\bm r},{\bm v},t).
 \end{equation}
 Now, we have to require $f_{l}({\bm r},{\bm v},t)$ to be a solution of the BL equation. This is easily seen to imply that $n$  does not depend on time and that it obeys the equation
 \begin{equation}
 \label{a1.13}
 \frac{\partial n({\bm r})}{\partial {\bm r}} = -\frac{1}{k_{B}T_{b}}\, \frac{\partial \phi}{\partial {\bm r}}.
 \end{equation}
 The solution of this equation is given by Eq.\ (\ref{2.20}), and then $f_{l}$  in Eq. (\ref{a1.11}) becomes $f_{st}$ in Eq.\, (\ref{2.22}).

 \end{document}